\documentclass[twocolumn,showpacs,preprintnumbers,nofootinbib]{revtex4}
\usepackage{amssymb}
\usepackage{amsmath}

\usepackage{graphicx}
\usepackage{dcolumn}
\usepackage{bm}

\draft
\input{tcilatex}

\begin{document}

\title{On the dynamical anomalies in the Hamiltonian Mean Field model}
\author{L. Velazquez}
\email{luisberis@geo.upr.edu.cu}
\affiliation{Departamento de F\'{\i}sica, Universidad de Pinar del R\'{\i}o, Mart\'{\i}
270, Esq. 27 de Noviembre, Pinar del R\'{\i}o, Cuba.}
\author{R. Sospedra}
\affiliation{Departamento de F\'{\i}sica Nuclear, Instituto Superior de Ciencias y
Tecnolog\'{\i}a Nucleares, Carlos III y Luaces, Plaza, La Habana, Cuba.}
\author{J.C. Castro}
\affiliation{Departamento de F\'{\i}sica, Universidad de Pinar del R\'{\i}o, Mart\'{\i}
270, Esq. 27 de Noviembre, Pinar del R\'{\i}o, Cuba.}
\author{F. Guzm\'{a}n}
\affiliation{Departamento de F\'{\i}sica Nuclear, Instituto Superior de Ciencias y
Tecnolog\'{\i}a Nucleares, Carlos III y Luaces, Plaza, La Habana, Cuba.}
\date{\today}

\begin{abstract}
We study the \textit{N}-dependence of the thermodynamical variables and the
dynamical behavior of the well-known \textbf{Hamiltonian Mean Field model}.
Microcanonical analysis revealed a thermodynamic limit which defers from the 
\textit{a priory} traditional assumption of the $N$-dependence of the
coupling constant $g$ as $\sim1/N$: to tend $N\rightarrow\infty$ keeping
constant $E/N^{3}$ and $g/N$, prescription which guarantees the extensivity
of the Boltzmann entropy. The analysis of dynamics leads to approximate the
time evolution of the magnetization density $\mathbf{m}$ by means of a
Langevin equation with multiplicative noise. This equation leads to a
Fokker-Planck's equation which is \textit{N}-independent when the time
variable is scaled by the \textit{N}-dependent time constant $\tau_{mac}=%
\sqrt{IN/g}$, which represents the characteristic time scale for the
dynamical evolution of the macroscopic observables derived from the
magnetization density. This results explains the origin of the slow
relaxation regimen observed in microcanonical numerical computations of
dynamics of this model system. Connection with the system self-similarity is
suggested.
\end{abstract}

\maketitle

\section{Introduction}

The present effort is devoted to study the \textit{N}-dependence in the
thermodynamical variables and dynamical behavior of the well-known the 
\textit{Hamiltonian mean field model} \cite{ant,lat1,lat2,lat3,lat4,zanette}%
. This model describes a system of \emph{N} planar classical spins
interacting through an infinite-range potential \cite{ant}:

\[
H_{N}\left( \theta,L;I,g\right) =K_{N}\left( L;I\right) +V_{N}\left(
\theta;g\right) 
\]%
\begin{equation}
\underset{i=1}{\overset{N}{=\sum}}\frac{1}{2I}L_{i}^{2}+\frac{1}{2}g\underset%
{i,j=1}{\overset{N}{\sum}}\left[ 1-\cos\left( \theta_{i}-\theta_{j}\right) %
\right] \text{,}   \label{Hamiltonian}
\end{equation}
where $\theta_{i}$ is the \emph{i-th} angle and $L_{i}$ the conjugate
variable representing the angular momentum, $I$ is the moment of inertia of
the rotator and $g$ the coupling constant. This is an inertial version of
the ferromagnetic X-Y model \cite{stan}, which interaction is not restricted
to first neighbors but is extended to all couples of spins.

Traditionally the coupling constant $g$ is suppose to be \emph{N}-dependent:

\begin{equation}
g=\frac{\gamma}{N},   \label{cc}
\end{equation}
assumption which makes $H$ formally extensive ($V\propto N$ when $%
N\rightarrow\infty$), since the energy remains non-additive and the system
can not be trivially divided in independent subsystems. A fundamental
observable of this model system is the magnetization density $\mathbf{m}$,
which is given by:

\begin{equation}
\mathbf{m}=\frac{1}{N}\underset{i=1}{\overset{N}{\sum}}\mathbf{m}_{i}\text{,}
\label{m}
\end{equation}
where $\mathbf{m}_{k}=\left[ \cos\theta_{k},\sin\theta_{k}\right] $ is the
intrinsic magnetization of the \emph{k}-th particle.

The canonical analysis of the model predicts a second-order phase transition
from a low-energy ferromagnetic phase with magnetization $m\approx1$ to a
high-energy paramagnetic phase with $m\approx0$. The dependence of the
energy density $U=E/N$ on the temperature $T$, the caloric curve, is given
by:

\begin{equation}
U=\frac{T}{2}+\frac{1}{2}\left[ 1-m^{2}\left( T\right) \right]   \label{te}
\end{equation}
($U$ and $T$ in units of $\gamma$) and this is shown at the Fig.(\ref{md}).
The critical point is at the energy $U_{c}=0.75$ corresponds to the critical
temperature of $T_{c}=0.5$ \cite{ant}.

The dynamical behavior of the HMF has been extensively investigated.
Microcanonical simulations of dynamics are in general in good agreement with
the canonical solution, except for a region below $U_{c}$, where some
anomalous phenomena have been observe, such as \textit{ensemble inequivalence%
}\emph{\ }and the \textit{slow relaxation regimen} towards the
Boltzmann-Gibbs' equilibrium \cite{lat2,lat3,lat4,zanette}.

This ensemble inequivalence, characterized by the existence of a negative
specific heat, appears as a dynamical feature. The origin of this anomalie
is explained by the existence of certain \emph{quasi-stationary state} \cite%
{lat3,lat4} which is not the canonical one and depends on the initial
conditions in the microcanonical numerical computation of the dynamics. In
this case, the system needs a long time to relax to the canonical
equilibrium state because of the duration of this metastable state has a
linear dependence with $N$. This fact has been interpreted by Tsallis as the
non-commutativity of the thermodynamic limit ($N\rightarrow\infty)$ with the
infinite time limit ( $t\rightarrow\infty$): when the first is performed
before the second, the system will not relax to the Boltzmann-Gibbs'
equilibrium.

Recent studies have shown that the slow-relaxation regime at long times is
clearly revealed by numerical realizations of the model, but \textit{no
traces of quasistationarity} are found during the earlier stages of the
evolution. This fact suggested the nonergodic properties of this system in
the short-time range, which could make the standard statistical description
unsuitable \cite{zanette}.

However, it there could be \textit{a mistake} in the understanding of the
macroscopic description of this model. There are many studies in long-range
Hamiltonian system where the validity of the thermodynamical description in
the thermodynamic limit is intimately related to the scaling with $N$ of the
thermodynamic variables and potentials. The problems with the nonextensive
nature of those systems are avoided using the Kac prescription \cite{kac} in
which the coupling constants are scaled by some power of $N$ in order to
deal with an extensive total energy $E$. It is well-known that this kind of
assumption \textit{affected the time scale} of the system evolution, so
that, its application should be taken with care.

\section{Microcanonical analysis}

M. Antoni, H. Hinrichsen and S. Ruffo carried out the microcanonical
analysis of the HMF in the ref.\cite{ant2}. However, in their work they
considered the equivalence of the microcanonical with the canonical ensemble
in the thermodynamic limit, which depends on the validity of the canonical
solution. In this section we will perform this analysis working directly on
the microcanonical ensemble.

The microcanonical states density $\Omega$ is calculated as follows:

\begin{equation}
\Omega\left( E,N;I,g\right) =\frac{1}{N!}\int\frac{d^{N}\theta d^{N}L}{%
\left( 2\pi\hslash\right) ^{N}}\delta\left[ E-H_{N}\left(
\theta,L;I,g\right) \right] \text{.}
\end{equation}
The integration by $d^{N}L$ yields:

\begin{equation}
\frac{1}{N!\Gamma\left( \frac{N}{2}\right) }\left( \frac{2\pi I}{\hslash^{2}}%
\right) ^{\frac{1}{2}N}\int\frac{d^{N}\theta}{\left( 2\pi\right) ^{N}}\left[
E-V_{N}\left( \theta;g\right) \right] ^{\frac {1}{2}N-1}\text{.}
\end{equation}
The microcanonical accessible volume $W$ is expressed as $W=\Omega g/2$%
\footnote{%
For continuous variables, $W$ is only well-defined after a coarsed grained
partition of phase space, which is the reason why it is considered an small
energy constant $\delta\varepsilon$ in order to make $W$ dimensionless.
However, during the thermodynamic limit $N\rightarrow\infty$, the selection
of $\delta\varepsilon$ does not matter if it is small.}. We are interested
in describing the large $N$ limit, where $W$ is given by:

\begin{equation}
W\simeq\left( \frac{2\pi Ig}{N\hslash^{2}}\right) ^{\frac{1}{2}N}\int d^{2}%
\mathbf{m}\left[ m^{2}+2U-1\right] ^{\frac{1}{2}N}f\left( \mathbf{m}%
;N\right) ,   \label{WW}
\end{equation}
where it was introduced the dimensionless parameter $U$:

\begin{equation}
E=gN^{2}U\text{,}
\end{equation}
and the function $f\left( \mathbf{m};N\right) $:

\begin{equation}
f\left( \mathbf{m};N\right) =\int\frac{d^{N}\theta}{\left( 2\pi\right) ^{N}}%
\delta\left[ \mathbf{m}-\frac{1}{N}\overset{N}{\underset{k=1}{\sum}}\mathbf{m%
}_{k}\right] \text{.}   \label{ff}
\end{equation}

We estimate the function $f\left( \mathbf{m};N\right) $ for large $N$ by
using the \emph{steepest decent method}. Expressing the delta function by
mean of the Fourier's integral representation, the Eq.(\ref{ff})\ is
rewritten as follows:

\[
f\left( \mathbf{m};N\right) =N^{2}\int \frac{d^{2}\mathbf{k}}{\left( 2\pi
\right) ^{2}}\exp \left[ iN\overline{\mathbf{k}}\cdot \mathbf{m}\right] %
\left[ I_{0}\left( -i\sqrt{\overline{\mathbf{k}}^{2}}\right) \right] ^{N}
\]%
where $\overline{\mathbf{k}}=\mathbf{k+}i\mathbf{x}$, $\mathbf{x\in R}^{2}$,
and $I_{0}\left( z\right) $ is the modified Bessel function of zero order.
It is easy to obtain the main contribution of the integral when $%
N\rightarrow \infty $ maximizing the integral function via the parameter $%
\mathbf{x}$. That is:

\[
f\left( \mathbf{m};N\right) \simeq N^{2}\exp\left\{ -N\left[ xm-\ln
I_{0}\left( x\right) \right] -\right. 
\]

\begin{equation}
\left. -\frac{1}{2}\ln\left[ \left( \frac{N}{2\pi}\right) ^{2}\frac{m\left(
x\right) }{x}\frac{d}{dx}m\left( x\right) \right] \right\} ,   \label{func}
\end{equation}
where $x=\left| \mathbf{x}\right| $ is related with $m=\left| \mathbf{m}%
\right| $ by:

\begin{equation}
m=m\left( x\right) =\frac{I_{1}\left( x\right) }{I_{0}\left( x\right) }\text{%
.}
\end{equation}
It is necessary to say that the validity of the steepest decent\ procedure
is ensured by the \textit{positivity} of the argument of the logarithmic
term in the Eq.(\ref{func}).

The main contribution to the entropy per particle $s\left( U,N;I,g\right)
=\ln W/N$ for $N$ large can be obtained by using again the steepest decent
method as follows:

\[
s_{o}\left( U,N;I,g\right) =\frac{1}{2}\ln\left( \frac{2\pi Ig}{N}\right) + 
\]

\begin{equation}
+\underset{x}{\max}\left\{ \frac{1}{2}\ln\left[ m^{2}\left( x\right) +\kappa%
\right] -xm\left( x\right) +\ln I_{0}\left( x\right) \right\} , 
\label{s per part}
\end{equation}
where $\kappa=2U-1$. The maximization yields:

\begin{equation}
\left[ \frac{m\left( x\right) }{x}-m^{2}\left( x\right) -\kappa\right] \frac{%
x}{m^{2}\left( x\right) +\kappa}\frac{d}{dx}m\left( x\right) =0\text{,} 
\label{con}
\end{equation}
which represents the energy dependence of magnetization density for the bulk
matter:

\begin{equation}
\begin{array}{c}
\begin{tabular}{lll}
$\left. 
\begin{array}{c}
m=m(x) \\ 
U=\frac{1}{2}m\left( x\right) /x+\frac{1}{2}\left[ 1-m^{2}\left( x\right) %
\right]%
\end{array}
\right\} $ & with & $x\in\left[ 0,\infty\right) $%
\end{tabular}
\\ 
\text{and }m=0\text{ \ if \ }U\geq U_{c}=0.75%
\end{array}
.
\end{equation}
The caloric curve (\ref{te}) is obtained from the canonical parameter $\beta$%
:

\begin{align}
\beta & =\frac{1}{T}=\frac{\partial s_{0}\left( U,N;I,g\right) }{\partial U}
\nonumber \\
& =\left\{ 
\begin{tabular}{lll}
$x/m\left( x\right) $ & with & $0\leq U<0.75$ \\ 
$1/\left( 2U-1\right) $ & if & $U\geq0.75$%
\end{tabular}
\ \right. .
\end{align}
This is the same result found by M. Antoni, H. Hinrichsen and S. Ruffo in
the ref.\cite{ant2}. Therefore, the subsequent analysis will be equivalent
to their work. The microcanonical analysis confirms the existence of a
second-order phase transition at the critical energy $U_{c}=0.75$ with a
critical temperature of $T_{c}=0.5$.

It is interesting to note that although the main term of the entropy per
particle is \textit{N}-independent, there is an additional contribution
which is \textit{N}-dependent and makes the Boltzmann entropy \textit{%
nonextensive} for an arbitrary \textit{N}-dependence of the coupling
constant $g$. Although the selection of the \textit{N}-behavior of the
coupling constant does not change the thermodynamical picture in the
microcanonical ensemble, it \textit{affect}s the extensive or nonextensive
character of the energy in the thermodynamic limit, as well as the temporal
scale in the dynamical behavior of this model system (see in the next
section).

In order to deal with an extensive Boltzmann entropy is necessary to assume
the following thermodynamic limit for the Hamiltonian mean field model:

\begin{equation}
N\rightarrow\infty,\text{~keeping constant }\frac{E}{N^{3}}\text{ and }\frac{%
g}{N}.   \label{thermodynamic limit}
\end{equation}
This assumtion defers from the one used by other investigators, where the 
\textit{N}-dependence of the coupling constant chosen in order to deal with
an extensive energy.

\FRAME{fhFU}{2.7665in}{2.3687in}{0pt}{\Qcb{Microcanonical description of the
HMF model in thermodynamic limit. Here it is clearly shown a second-order
phase transition at $U_{c}=0.75$, where $T_{c}=0.5$ and $m=0$.}}{\Qlb{md}}{%
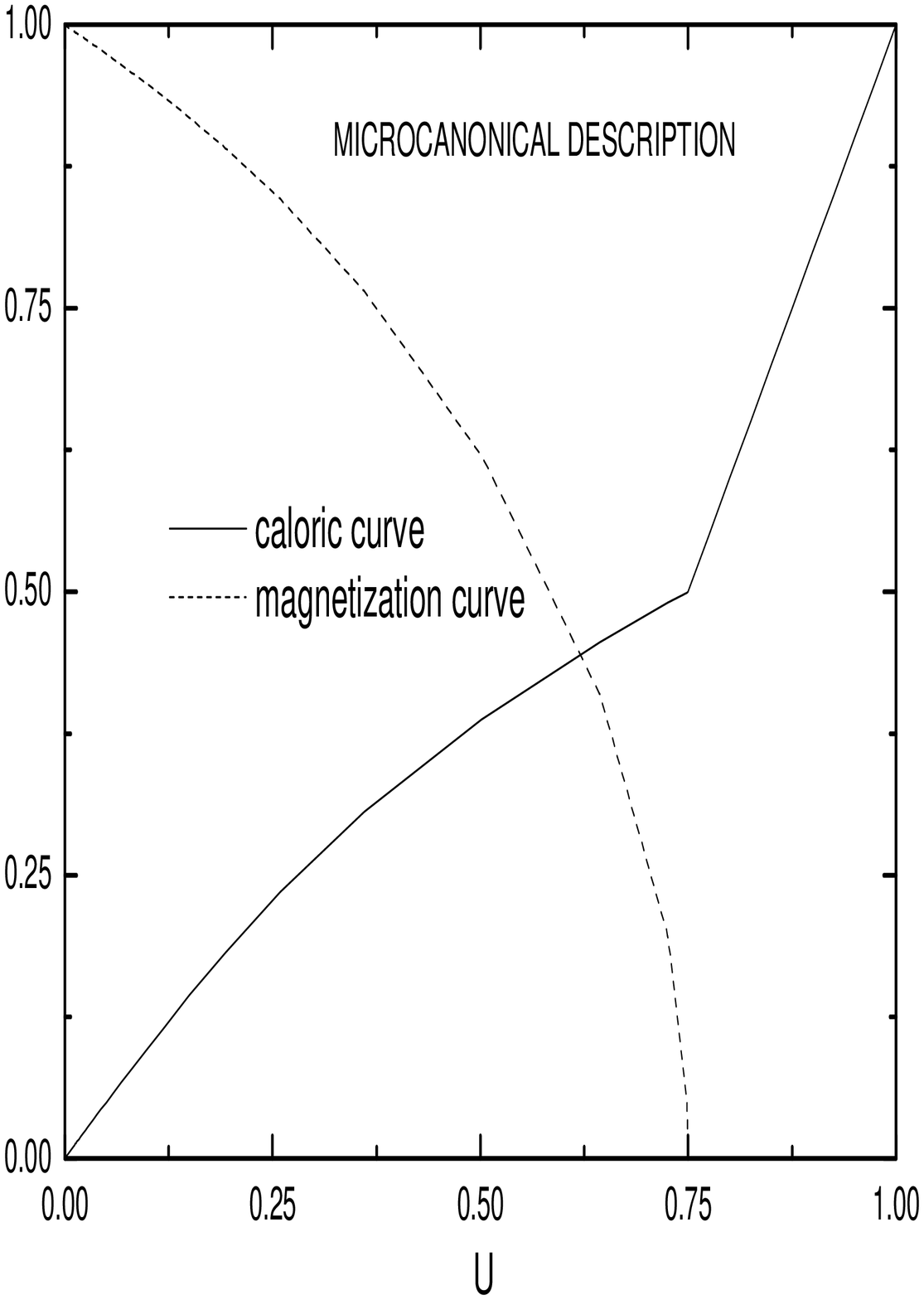}{\special{ language "Scientific Word"; type "GRAPHIC"; display
"USEDEF"; valid_file "F"; width 2.7665in; height 2.3687in; depth 0pt;
original-width 6.2258in; original-height 8.8047in; cropleft "0"; croptop
"1"; cropright "1"; cropbottom "0"; filename 'md.eps';file-properties
"XNPEU";}}

\section{The origin of the slow relaxation regimen}

According to the Hamiltonian of the model, its motion equations are given by:

\begin{align}
\dot{\theta}_{k} & =\frac{1}{I}L_{k}\text{,}  \nonumber \\
\dot{L}_{k} & =gN\left( m_{y}\cos\theta_{k}-m_{x}\sin\theta_{k}\right) \text{%
,}   \label{motion equations}
\end{align}
where $k=1,2,...N$ and it was used the expression given in Eq.(\ref{m}). It
is convenient to introduce the following characteristic units for time and
momentum:

\begin{equation}
\tau_{0}=\sqrt{\frac{I}{gN}}\text{ and }L_{0}=\sqrt{IgN}. 
\label{characteristics units}
\end{equation}
Note that, according to the thermodynamic limit (\ref{thermodynamic limit}),
the characteristic time $\tau_{0}$ decreases with the $N$ increasing as $%
\tau_{0}\propto1/N$, while the characteristic momentum $L_{0}$ increases as $%
L_{0}\propto N$. However, when it is assumed the \emph{N}-dependence (\ref%
{cc}) for the coupling constant, the characteristic time $\tau_{0}$ is \emph{%
N}-independent. The time scale of $\tau_{0}$ gives the characteristic time
scale for the evolution of each rotator, so that, this time units is
relevant to microscopic level. In the numerical computation of the dynamics
this characteristic time scale should be taken into account in order to
ensure a great accuracy of the energy conservation.

Using the above characteristic units, as well as introducing the
tridimensional vectors $\mathbf{m}_{k}=\left( \cos\theta_{k},\sin\theta
_{k},0\right) $ and $\mathbf{o}_{k}=\left( 0,0,L_{k}\right) $, the motion
equations (\ref{motion equations}) are rephrased as follows:

\begin{equation}
\mathbf{\dot{m}}_{k}=\mathbf{o}_{k}\times\mathbf{m}_{k}\text{, }\mathbf{\dot{%
o}}_{k}=\mathbf{m}_{k}\times\mathbf{m}\text{.}   \label{vectorial form}
\end{equation}
The magnetization density obeys the following dynamical equation:

\begin{equation}
\mathbf{\dot{m}}=\mathbf{\Upsilon}=\mathbf{\Omega}\times\mathbf{m}+\Lambda%
\mathbf{m},   \label{magnetization dynamics}
\end{equation}
where $\mathbf{\Upsilon}$ is the velocity of change of the magnetization
density, $\mathbf{\Omega}$, the angular velocity which characterizes the
dynamical behavior of the $\mathbf{m}$ orientation, while $\Lambda$
characterizes the behavior of its modulus. These quantities are given by:

\begin{equation}
\mathbf{\Upsilon=}\frac{1}{N}\overset{N}{\underset{k=1}{\sum}}\mathbf{o}%
_{k}\times\mathbf{m}_{k},   \label{mag vel}
\end{equation}

\begin{equation}
\mathbf{\Omega=}\frac{1}{Nm}\overset{N}{\underset{k=1}{\sum}}\left( \mathbf{m%
}_{k}\cdot\mathbf{m}\right) \mathbf{o}_{k}\mathbf{,~}\Lambda \mathbf{=}\frac{%
1}{Nm}\overset{N}{\underset{k=1}{\sum}}\mathbf{o}_{k}\times\mathbf{m}%
_{k}\cdot\mathbf{m.}
\end{equation}

Since $\mathbf{o}_{k}\times\mathbf{m}_{k}$'s do not have a definite sign,
the magnetization velocity of change $\mathbf{\Upsilon}$ is a fast
fluctuating quantity with a vanishing short time average (a time average
along temporal interval comparable with the characteristic time of rotators
evolution $\tau_{0}$) and standard deviation $\sigma_{\mathbf{\Upsilon}}$ $%
\thicksim 1/\sqrt{N}$. This fact provokes that the short time average of the
magnetization $\mathbf{m}$ varies slowly in comparison with the
characteristic time scale of rotators evolution. Thus, there is two time
scales in the dynamical evolution of the system: the microscopic time scale, 
$\tau _{mic}=\tau_{0}$, which is given by the characteristic time of the
rotators evolution, and the macroscopic time scale, $\tau_{mac}$, which is
given by the characteristic time of the short time average of magnetization
density evolution, being this last considerably greater than the first time
scale, $\tau_{mac}\gg\tau_{mic}\,$.

We are interested in studying the dynamical evolution of the macroscopic
observables for temporal scale comparable with $\tau_{mac}$. Since $\tau
_{mac}\gg\tau_{mic}$, at first approximation the dynamical evolution of $%
\mathbf{m}$ could be modeled by a Langevin equation with \textit{%
multiplicative noise} \cite{celia,foc}:

\begin{equation}
\mathbf{\dot{m}}=\mathbf{\Upsilon=}\sigma_{1}\left( m;U\right) \xi
_{1}\left( t\right) \mathbf{e}_{1}+\sigma_{2}\left( m;U\right) \xi
_{2}\left( t\right) \mathbf{e}_{2}.   \label{langevin}
\end{equation}
whose parameters $\sigma_{1}\left( m;U\right) $ and $\sigma_{2}\left(
m;U\right) $ could be estimated by using a microcanonical average with fixed
magnetization density. This last estimation is a reasonable approximation
since the magnetization density only exhibits very small fluctuations around
its short time average for $N~$large along time intervals comparable with $%
\tau_{mic}$. Here, $\mathbf{e}_{1}$ and $\mathbf{e}_{2}$ are unitary vectors
in the parallel and perpendicular directions of $\mathbf{m}$ respectively,
being $\mathbf{e}_{3}=\mathbf{e}_{1}\times\mathbf{e}_{2}\equiv\left(
0,0,1\right) $, the unitary vector along the third axis. $\xi_{1}\left(
t\right) $ and $\xi_{2}\left( t\right) $ are Gaussian processes:

\begin{equation}
\left\langle \xi_{i}\left( t\right) \xi_{j}\left( t^{\prime}\right)
\right\rangle =\tau_{i}\delta_{ij}\delta\left( t-t^{\prime}\right)
\end{equation}
being $\tau_{i}$ the correlation times. The presence of $\delta_{ij}$ is due
to the vanishing of the microcanonical average of $\left\langle \mathbf{%
\Upsilon}_{1}\mathbf{\Upsilon}_{2}\right\rangle =\left\langle \mathbf{%
\Upsilon\cdot e}_{1}\mathbf{\Upsilon\cdot e}_{2}\right\rangle $ with fixed
magnetization density, which is easily proved by mean of a procedure
analogue to the one used in the previous section for the determination of
Boltzmann entropy. As already mentioned, the standard deviations of the
components of the vector $\mathbf{\Upsilon}$ depend on $N$ as $\sigma
_{j}\left( m;U\right) \thicksim1/\sqrt{N}$.

It is not difficult to understand that the correlation times $\tau_{i}$ are
comparable with the microscopic time scale $\tau_{mic}$: the correlation
times possess the same temporal scale of the characteristic time where the
differenciable character of $\mathbf{\Upsilon}$ is perceived, which is
precisely characteristic time scale of rotators evolution.

According to the general theory, to this equation is associated the
Fokker-Planck equation of the form:

\begin{equation}
\frac{\partial}{\partial t}F=\partial_{m^{\alpha}}\left\{ \frac{1}{2}\tau
_{j}\sigma_{j}e_{j}^{\alpha}\partial_{m^{\beta}}\left[ \sigma_{j}e_{j}^{%
\beta}F\right] \right\} ,   \label{Fokker-Planck}
\end{equation}
which describes the dynamics of the magnetization density distribution
function $F=F\left( m,t;U\right) $. In this expression, the $e_{j}^{\alpha}$%
's represent the components of $\mathbf{e}_{j}$. Taking into account the 
\textit{N}-dependence of $\sigma_{j}$, the right hand of the equation (\ref%
{Fokker-Planck}) \textit{decreases} as $1/N$ during the $N$ increasing.
Scaling the time scale as $t\rightarrow t=Nt^{\prime}$, the Fokker-Planck
equation becomes \textit{N}-independent. This means that the characteristic
time scale for the dynamical evolution of $F\left( m,t;U\right) $, as well
as all those macroscopic observables derived from $\mathbf{m}$, \textit{have
a linear dependence on} $N$ during the increasing of system size (in $%
\tau_{0}$ units).

This is precisely the result which was observed in the numerical computation
of the dynamics, where the relaxation time for the dynamical temperature $%
T_{D}\left( t\right) =2K\left( t\right) /gN^{2}=2U-1+m^{2}\left( t\right) $
grows proportional to $N$ with the $N$ increasing. As already mentioned, the
time scale of such dynamical behaviors is precisely the macroscopic time
scale $\tau_{mac}$:

\begin{equation}
\tau_{mac}\sim\tau_{0}N=\sqrt{\frac{IN}{g}}.
\end{equation}

Thus, when the thermodynamil limit (\ref{thermodynamic limit}) is taken into
account, the relaxation time is finite when $N$ is tended to infinity.
Therefore, \textit{no anomalous dynamical behavior will be observed}: the
thermodynamic limit ($N\rightarrow\infty$) necesary for the ensemble
equivalence between the microcanonical ensemble with the canonical one, will
commute with the infinite time limit ($t\rightarrow\infty$) necesary for the
equilibration of the temporal average of the physical observables. Thus, the
slow relaxation regimen observed in the microcanonical numerical computation
of dynamics appears as consequence of the consideration of the \textit{N}%
-dependence of the coupling constant (\ref{cc}). The above analysis shows
that the \textit{N}-dependence of the coupling constant which ensures the
extensivity of the Boltzmann entropy is more appropriate than the one which
guarantees the extensivity of the energy. The first allows us to carry out a
well-defined macroscopic description for the Hamiltonian mean field model,
instead, the second leads to the existence of dynamical anomalies.

\section{Concluding remarks}

As already showed, the microcanonical analysis revealed a thermodynamic
limit (\ref{thermodynamic limit}) which defers from the traditional
assumption of the $N$-dependence of the coupling constant (\ref{cc}). This
thermodynamic limit is intimately related with the existence of the
following self-similarity scaling behavior of the thermodynamical variables
of the Hamiltonian mean field model:

\begin{equation}
\left. 
\begin{array}{c}
N\rightarrow N\left( \alpha \right) =\alpha N_{,} \\ 
E\rightarrow E\left( \alpha \right) =\alpha ^{3}E, \\ 
g\rightarrow g\left( \alpha \right) =\alpha g, \\ 
I\rightarrow I\left( \alpha \right) =I,%
\end{array}%
\right\} \Rightarrow W\left( \alpha \right) =\mathcal{F}\left[ W\left(
1\right) ,\alpha \right]   \label{scaling laws}
\end{equation}%
where the functional $\mathcal{F}\left[ W,\alpha \right] $ defines an
exponential self-similarity scaling laws for the microcanonical volume $W$:

\begin{equation}
\mathcal{F}\left[ W,\alpha\right] =\exp\left[ \alpha\ln W\right] ,
\end{equation}
being

\begin{equation}
W\left( \alpha\right) =W\left[ E\left( \alpha\right) ,N\left( \alpha\right)
;g\left( \alpha\right) ,I\left( \alpha\right) \right] ,
\end{equation}
where $\mathcal{F}\left[ W,\alpha\right] $ satisfies the self-similarity
condition:

\begin{equation}
\mathcal{F}\left[ \mathcal{F}\left[ W,\alpha_{1}\right] ,\alpha_{2}\right] =%
\mathcal{F}\left[ W,\alpha_{1}\alpha_{2}\right] .
\end{equation}

The reader may understand that this self-similarity scaling behavior is a
generalization of the extensive properties exhibited by the traditional
systems. As already discussed in our previous works, this self-similarity
properties could be crucial in order to develop a well-defined
thermodynamical description for a Hamiltonian nonextensive system \cite%
{vel1,vel2,vel3}. We showed in the ref. \cite{vel3} how self-similarity can
be used in order to analyze the necessary conditions for the validity of
Tsallis' Nonextensive Statistics \cite{tsal1}. The results obtained in the
present paper seem to be consistent with this idea too.

In future works we hope to obtain an analytical expression for the
parameters appearing in the Fokker-Planck equation (\ref{Fokker-Planck}) in
order to carry out a comparative study of this equation with the numerical
computation of dynamics, since in this work we only concentrate on the
analysis of their \textit{N}-dependence.

\end{document}